\documentclass[
]{ws-procs9x6}

\usepackage{xspace} 
\usepackage{units}  
\usepackage[absolute]{textpos}

\newcommand{\CNNcan}{CNN$ \text{can}$\xspace}
\newcommand{\CNNlr}{CNN$ \text{lr}$\xspace}
\newcommand{\CNNran}{CNN$ \text{ran}$\xspace}
\newcommand{\symbBias}{\ensuremath{z}\xspace}
\newcommand{\symbDelay}{\ensuremath{\tau}\xspace}
\newcommand{\symbDimensionM}{\ensuremath{d_e}\xspace}
\newcommand{\symbDimX}{\ensuremath{\mathcal M}\xspace}
\newcommand{\symbDimY}{\ensuremath{\mathcal N}\xspace}
\newcommand{\symbGlobFehler}{\ensuremath{E_g}\xspace}
\newcommand{\symbInput}{\ensuremath{u}\xspace}
\newcommand{\symbInputBild}{\ensuremath{U}\xspace}
\newcommand{\symbInterdepmass}{\ensuremath{S^{(k)}}\xspace}
\newcommand{\symbInterdepRichtung}{\ensuremath{ S^{(k)}_\text{asymm}}\xspace}
\newcommand{\symbKopplungFB}{\ensuremath{\mathcal A}\xspace}
\newcommand{\symbKopplungFF}{\ensuremath{\mathcal B}\xspace}
\newcommand{\symbNachbarn}{\ensuremath{k}\xspace}
\newcommand{\symbOutput}{\ensuremath{y}\xspace}
\newcommand{\symbOutputBild}{\ensuremath{Y}\xspace}
\newcommand{\symbRefBild}{\ensuremath{Y^\text{ref}_v}\xspace}
\newcommand{\symbSphaere}{\ensuremath{\mathcal I}\xspace}
\newcommand{\symbState}{\ensuremath{x}\xspace}
\newcommand{\symbStateBild}{\ensuremath{X}\xspace}
\newcommand{\symbSystemA}{\ensuremath{A}\xspace}
\newcommand{\symbSystemB}{\ensuremath{B}\xspace}
\newcommand{\symbTestElem}{\ensuremath{v}\xspace}
\newcommand{\symbTestSet}{\ensuremath{V}\xspace}
\newcommand{\symbTtrans}{\ensuremath{\tau_\text{trans}}\xspace}
\newcommand{\symbZustandA}{\ensuremath{a}\xspace}
\newcommand{\symbZustandB}{\ensuremath{b}\xspace}

\renewcommand{\labelitemi}{$\rule[0.5ex]{0.3em}{0.3em}\medspace$}

\begin{document}

\title{MEASURING DIRECTED INTERACTIONS USING CELLULAR NEURAL NETWORKS WITH COMPLEX CONNECTION TOPOLOGIES}

\author{H. DICKTEN\textsuperscript{1,2,3,*}, C. E. ELGER\textsuperscript{1}, K. LEHNERTZ\textsuperscript{1,2,3}}

\address{
    \textsuperscript{1}Department of Epileptology, University of Bonn,\\ Sigmund-Freud-Stra{\ss}e~25, 53105~Bonn, Germany\\
    \textsuperscript{2}Helmholtz-Institute for Radiation and Nuclear Physics, University of Bonn,\\ Nussallee~14--16, 53115~Bonn, Germany\\
    \textsuperscript{3}Interdisciplinary Center for Complex Systems, University of Bonn,\\ Br{\"u}hler Stra{\ss}e~7, 53175~Bonn, Germany\\
    \textsuperscript{*}E-mail: \texttt{hdickten@uni-bonn.de}
        }

\begin{abstract}
We advance our approach of analyzing the dynamics of interacting complex systems with the nonlinear dynamics of interacting nonlinear elements.
We replace the widely used lattice-like connection topology of cellular neural networks (CNN) by complex topologies that include both short- and long-ranged connections.
With an exemplary time-resolved analysis of asymmetric nonlinear interdependences between the seizure generating area and its immediate surrounding we provide first evidence for complex CNN connection topologies to allow for a faster network optimization together with an improved approximation accuracy of directed interactions.
\end{abstract}

\keywords{CNN; Directed Interactions; Nonlinear Interdependence; iEEG; Complex Networks; Time Series Analysis; Seizure Prediction Device}

\begin{textblock*}{14cm}(3cm,27cm)
    \noindent R. Tetzlaff and C. E. Elger and K. Lehnertz (2013), \emph{Recent Advances in Predicting and Preventing Epileptic Seizures}, page 242--252, Singapore, World Scientific.\\
    Copyright 2013 by World Scientific.
\end{textblock*}

\bodymatter
\section{Introduction}
Synchronization phenomena play an important role in nearly all fields of science, including physics, chemistry, economy, and the neurosciences~\cite{Pikovsky2001, Boccaletti2002}.
The human epileptic brain can be regarded as a prominent example in which different forms of synchronization can be observed.
Estimators for synchronization~\cite{Lehnertz2009b, Lehnertz2011} are highly attractive to characterize interactions between brain areas involved in ictogenesis.

Promising computational platforms for approximating these estimators are, among other approaches\cite{Abdelhalim2011}, Cellular Neural (or Nonlinear) Networks (CNN) as they are capable of universal computation and offer massive computing power while minimizing space and energy consumption and are already available as analogue integrated circuits~\cite{Chua2002, Chernihovskyi2005, Tetzlaff2006, Roska2008}.
Recent studies have shown that the approach of analyzing the dynamics of interacting complex systems with the nonlinear dynamics of interacting nonlinear elements can also be extended to the concepts of phase synchronization\cite{Sowa2005} and generalized synchronization\cite{Krug2007}.
With the latter concept symmetric and asymmetric nonlinear interdependence measures can be defined that allow one to characterize strength and direction of interactions~\cite{Arnhold1999, Krug2007,Chicharro2009,Andrzejak2011,Lehnertz2011b}.

We investigated whether a CNN-based characterization of directed interactions can further be improved by modifying the canonical Chua--Yang CNN~\cite{Chua1988a}.
This CNN consists of a regular (lattice-like) arrangement of cells which we replace by complex topologies~\cite{Watts1998, Tsuruta2003}.
We evaluate approximation accuracies through the analysis of directed interactions in long-term, multi-channel, intracranial electroencephalographic (iEEG) recordings from an epilepsy patient.

\section{Methods}
\subsection{Nonlinear Interdependencies}
Let \symbSystemA and \symbSystemB denote two dynamical systems and let $\symbZustandA_n, n=1, \ldots, N$ and $\symbZustandB_n, n=1, \ldots, N$ denote time series of some observable of the respective system.
With
\begin{align}
    \vec{\symbZustandA}_n = \bigl( \symbZustandA_n, \ldots \symbZustandA_{n-(\symbDimensionM-1)\symbDelay}\bigr)
    \quad\text{and}\quad
    \vec{\symbZustandB}_n = \bigl( \symbZustandB_n, \ldots,\symbZustandB_{n-(\symbDimensionM-1)\symbDelay}\bigr)
\end{align}
we denote the reconstructed delay vectors in state space~\cite{Whitney1936, Takens1981} with an appropriate chosen time delay \symbDelay and embedding dimension \symbDimensionM.
Given some reference point in state space, the {mean-squared Euclidean distance} to its \symbNachbarn nearest neighbors reads:
\begin{equation}
    R_n^{( \symbNachbarn )} (\symbSystemA) = \frac{1}{\symbNachbarn } \sum_{j=1}^{\symbNachbarn}  \left( \vec {\symbZustandA}_n - \vec {\symbZustandA}_{r_{n,j}}\right)^2,
    \label{hd:eq:euklidAbstandZR}
\end{equation}
where $r_{n,j}$, $j=1,\ldots,\symbNachbarn$ denote the time indices of the \symbNachbarn nearest neighbors of $\vec{\symbZustandA}_n$.
With $s_{n,j}$, $j=1,\ldots,\symbNachbarn$ as time indices of the \symbNachbarn nearest neighbors of $\vec{\symbZustandB}_n$,
$R_n^{( \symbNachbarn )} (\symbSystemB)$ is defined analogously.
In addition, the \emph{\symbSystemB-conditioned} mean-squared Euclidean distance in the state space of system \symbSystemA is derived by replacing the nearest neighbors of $\vec{\symbZustandA}_n$ by the equal-time partners of the nearest neighbors of $\vec{\symbZustandB}_n$:
\begin{equation}
    R_n^{( \symbNachbarn )} (\symbSystemA|\symbSystemB) = \frac{1}{\symbNachbarn } \sum_{j=1}^{\symbNachbarn } \left( \vec {\symbZustandA}_n - \vec {\symbZustandA}_{s_{n,j}}\right)^2.
    \label{hd:eq:kondAbstandZR}
\end{equation}
$R_n^{( \symbNachbarn )} (\symbSystemB|\symbSystemA)$ is defined in complete analogy, and the
nonlinear interdependence \symbInterdepmass then reads~\cite{Arnhold1999}
\begin{align}
    S^{( \symbNachbarn )} (\symbSystemA|\symbSystemB) = \frac{1}{M} \sum^{M}_{n=1} \frac{R_n^{( \symbNachbarn )} (\symbSystemA)}{R_n^{( \symbNachbarn )} (\symbSystemA|\symbSystemB)},
    \label{hd:eq:interdepSglobal}
\end{align}
where $M$ denotes the total number of state space vectors.
Strength and direction of interactions can be characterized via a symmetric and asymmetric measure:
\begin{align}
    \begin{split}
        S^{( \symbNachbarn )}_{\text{symm}} &=  \frac{S^{( \symbNachbarn )} (\symbSystemA|\symbSystemB) + S^{( \symbNachbarn )} (\symbSystemB|\symbSystemA)}{2}\\
        S^{( \symbNachbarn )}_{\text{asymm}} &= \frac{S^{( \symbNachbarn )} (\symbSystemA|\symbSystemB) - S^{( \symbNachbarn )} (\symbSystemB|\symbSystemA)}{2}.
    \end{split}
    \label{hd:eq:interdepSymmetrieS}
\end{align}

\subsection{Cellular Neural Networks (CNN)}
Artificial Neural Networks (ANNs) are computational tools inspired by the brain that have found extensive utilization in complex real-world problems~\cite{Priddy2005}.
An ANN consists of simple artificial \textit{cells} or \textit{processing units} which are connected via \textit{edges}, but there exists no single formal definition of what an ANN exactly is.
ANNs feature characteristics such as high parallelism, intrinsic nonlinearity, as well as fault and noise tolerance.
More importantly, ANNs can be trained using a set of given examples and offer the ability to generalize~\cite{Watkin1993, HaykinBook1999}.

A Cellular Neural Network (CNN) is a subset of ANN where---in contrast to a Hopfield network\cite{Hopfield1982}---only local connections between cells are allowed\cite{Chua1988a}.
Hence the number of edges increases only linearly with the number of cells of the network.

\subsubsection{Dynamics of a CNN}\label{hd:sec:dynamic}
\begin{figure}
 \centering
 \includegraphics[width=0.5\columnwidth]{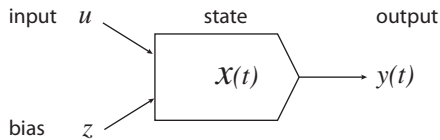}
  \caption{Representation of a single cell of a CNN. Each cell is a dynamical system with state $\symbState(t)$ evolving according to some state equation.}
 \label{hd:fig:zelle}
\end{figure}

Following Ref.~\refcite{Chua1998}, a CNN is a spatial arrangement of locally connected cells, where each cell is a dynamical system which has an input \symbInput, bias \symbBias, output $\symbOutput(t)$ and state $\symbState(t)$ evolving according to some state equation (cf. \fref{hd:fig:zelle}).

Let us first consider a CNN which consists of a two-dimensional $\symbDimX\times\symbDimY$ translation-invariant lattice of cells with nonlinear interactions.
The corresponding state equation for cell $(i,j)$ ($i \in [1, \symbDimX]$ and $j \in [1, \symbDimY]$) reads
\begin{align}
	\dot{\symbState}_{ij}(t) = - {\symbState}_{ij}(t) + \sum_{lm \in {\symbSphaere}_{ij}} \!\!{\symbKopplungFB}_{lm} \bigl({\symbOutput}_{lm}(t)\bigr) + \sum_{lm \in {\symbSphaere}_{ij}} \!\!{\symbKopplungFF}_{lm} \bigl({\symbInput}_{lm}\bigr)  + {\symbBias},
	\label{hd:eq:cnnZustandsgleichung}
\end{align}
where $\symbSphaere_{ij}$ denotes the \emph{sphere of influence}
of cell $(i,j)$, and $l,m \in \symbSphaere_{ij}$. ${\symbKopplungFB}_{lm}$ and ${\symbKopplungFF}_{lm}$ denote the feedback and feed-forward template functions, respectively.
In order to present the time series $\symbZustandA_n, \symbZustandB_n, N=4096$ to the network, we used a line wise alignment, i.e., the rightmost cell in a row is connected to the leftmost cell in the following row. Time series $\symbZustandA_n$ was assigned to the input \symbInput and time series $\symbZustandB_n$ to the initial state $\symbState(0)$ of the CNN.
Together with the chosen boundary condition this alignment preserves the temporal order of the time series.
Nevertheless it may introduce correlations between uncorrelated data points within the time series.

\begin{figure}
    \centering
    \begin{minipage}[t]{0.3\columnwidth}
        \includegraphics[width=\columnwidth]{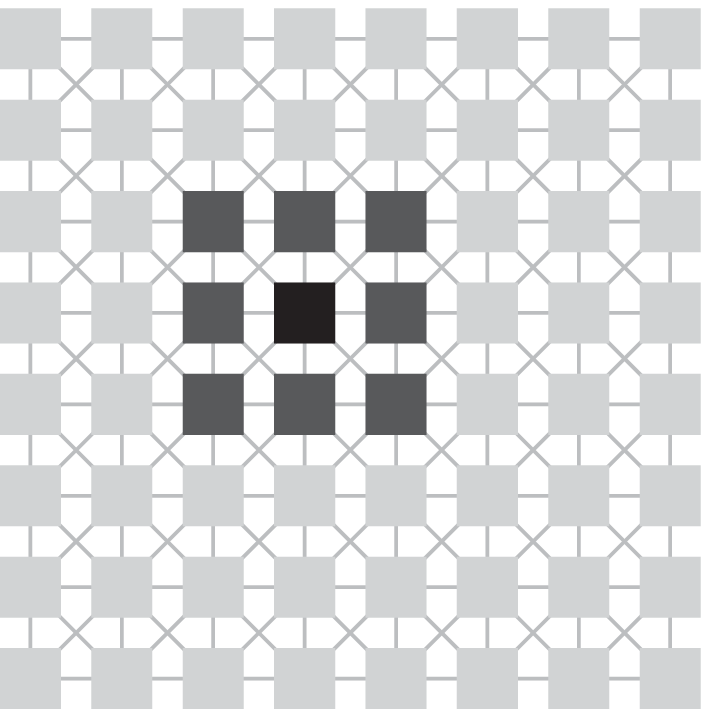}\\
        \footnotesize(a) \CNNcan
    \end{minipage}
    \hfill
    \begin{minipage}[t]{0.3\columnwidth}
        \includegraphics[width=\columnwidth]{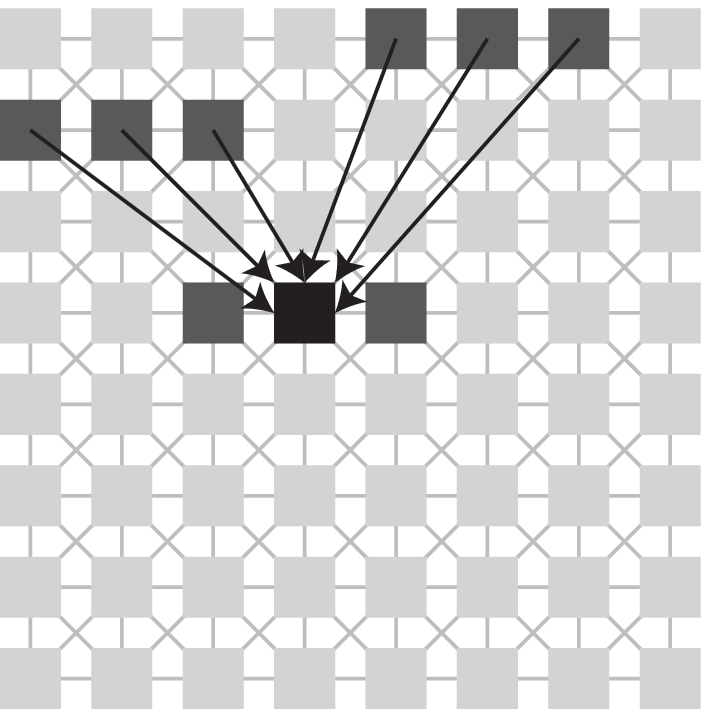}\\
        \footnotesize(b) \CNNlr
    \end{minipage}
    \hfill
    \begin{minipage}[t]{0.3\columnwidth}
        \includegraphics[width=\columnwidth]{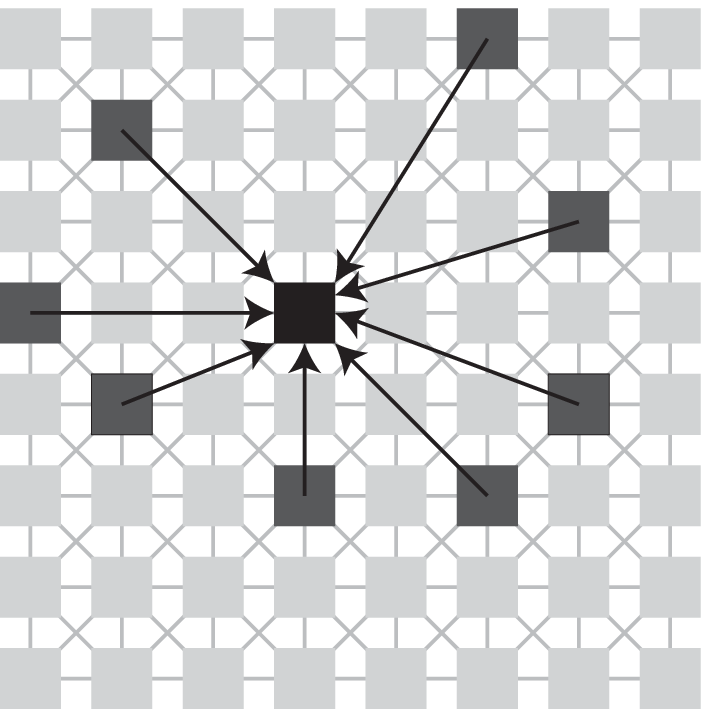}\\
        \footnotesize(c) \CNNran
    \end{minipage}
    \caption{CNN with the canonical (a) and with complex connection topologies (b and c). In (b) long-ranged connections are chosen according to the time of the first maximum of the autocorrelation function for each time series separately. In (c) short- and long-ranged connections are chosen at random.}
    \label{hd:fig:topos}
\end{figure}

Following Ref.~\refcite{Krug2007}, we here define the canonical Chua--Yang \CNNcan as a quadratic network arrangement ($\symbDimX \times \symbDimY = 64$) with a minimum possible $3 \times 3$ sphere of influence as well as polynomial-type template functions of order three (cf. \fref{hd:fig:topos} left).
In order to investigate whether complex connection topologies allow for an improved CNN-based characterization of directed interactions, we additionally consider two modified versions of \CNNcan:

With \CNNlr we define a topology, in which some short- and long-ranged connections between cell $(i,j)$ and cells of its sphere of influence are introduced.
The distance of the long-ranged connections corresponds to the time of the first maximum of the autocorrelation function of time series $\symbZustandA_n$ and $\symbZustandB_n$, respectively (normalized by the sampling interval), thus minimizing the aforementioned effect of connecting possibly uncorrelated data. Short-ranged connections exist between cells $(i,j)$ and $(i,j-1)$ and between cells $(i,j)$ and $(i,j+1)$ (cf. \fref{hd:fig:topos} middle).
Eventually we fully relax the canonical topology by choosing connections between cell $(i,j)$ and eight other cells (that comprise the sphere of influence) at random (\CNNran; cf. \fref{hd:fig:topos} right).
Each sphere of influence remained translation-invariant and consisted of nine (eight to other cells and one to itself) connections to ensure comparability between different topologies.

\subsubsection{Optimization and Validation}\label{hd:sec:optimization}
In the following, we denote with \symbInputBild, $\symbStateBild(t)$, and $\symbOutputBild(t)$ the inputs, states and outputs of all CNN cells.
In order to optimize the networks we used an evolutionary algorithm\cite{Holland1975, Kunz2000b} with the following parameters: population size: 50, number of survivors: 10, number of immigrants: 10, maximum number of iteration steps: 300.

We performed an in-sample optimization of our CNN using $\symbTestSet=20$ representative pairs of time series of \unit[20.48]{s} duration (in total $\approx$ \unit[7]{min} EEG) each along with the corresponding value of the nonlinear interdependence measure calculated according to \eref{hd:eq:interdepSymmetrieS} (denoted as $S^\text{ref.}$ in the following) with $\symbDelay = 5, \symbDimensionM = 10$, and $\symbNachbarn = 6$.
Half of the values represented weakly dependent time series ($S^\text{ref. low}_\text{asymm.} \in [-0.04, -0.02]$), and the other half stronger dependent time series ($S^\text{ref. high}_\text{asymm.} \in [0.04, 0.06]$), respectively~\cite{Krug2008b}.
To check for possible over-optimization of our CNN, we performed an additional out-of-sample validation using a similar setup as before but with another set of 20 pairs of time series along with the corresponding values of the nonlinear interdependence.

The approximated asymmetric nonlinear interdependence measure \symbInterdepRichtung was obtained by the rescaled mean output of all cells:
\begin{equation}
    S_\text{asymm.}^\text{CNN} = \left( \frac{S^\text{ref. high}_\text{asymm.} - S^\text{ref. low}_\text{asymm.}}{\symbDimX \symbDimY} \sum^{\symbDimX, \symbDimY}_{i,j=0} \frac{\symbOutput_{i,j}(\symbTtrans) + 1}{2}\right) + S^\text{ref. low}_\text{asymm.}
    \label{hd:eq:approxinterdep}
\end{equation}
After choosing random initial values for the components of templates \symbKopplungFB and \symbKopplungFF and for the global cell bias \symbBias the global error was minimized:

\begin{equation}
    \symbGlobFehler = \frac{1}{\symbTestSet}\sum^{\symbTestSet-1}_{\symbTestElem=0}\left( \frac{1}{4 \symbDimX\symbDimY} \sum^{\symbDimX, \symbDimY}_{i,j=0} \left(\symbOutput_{i,j,\symbTestElem}(\symbTtrans) - \symbRefBild \right)^2\right)
    \label{hd:eq:globFehler}
\end{equation}
where \symbTtrans denotes some fixed transition time\cite{Kunz2000, Krug2006}. All calculations were performed using our distributed computing system~\cite{Muller2006}, and network simulations were performed with Conedy~\cite{Rothkegel2012}.

\section{CNN-based iEEG Analysis}
We analyzed directed interactions in multi-channel iEEG recordings from an epilepsy patient who underwent presurgical evaluation of a left-sided mesial temporal lobe epilepsy.
After selective amygdalo-hippocampectomy the patient is completely seizure-free.
The patient had signed informed consent that the clinical data might be used and published for research purposes. The study protocol had previously been approved by the ethics committee of the University of Bonn. iEEG was measured from bilaterally implanted intrahippocampal depth electrodes. iEEG data were sampled at \unit[200]{Hz} using a \unit[16]{bit} analog-to-digital converter and filtered within a frequency band of \unit[0.5--85]{Hz}.

In the following we report our findings of estimating directed interactions between the seizure generating area and its immediate surrounding in a time-resolved manner (moving window analysis; non-overlapping windows of \unit[20.48]{s} duration). We here restrict ourselves to an interictal recording lasting for about 25~hours.
\begin{figure}[ht]
    \centering
    \includegraphics[width=0.9\columnwidth]{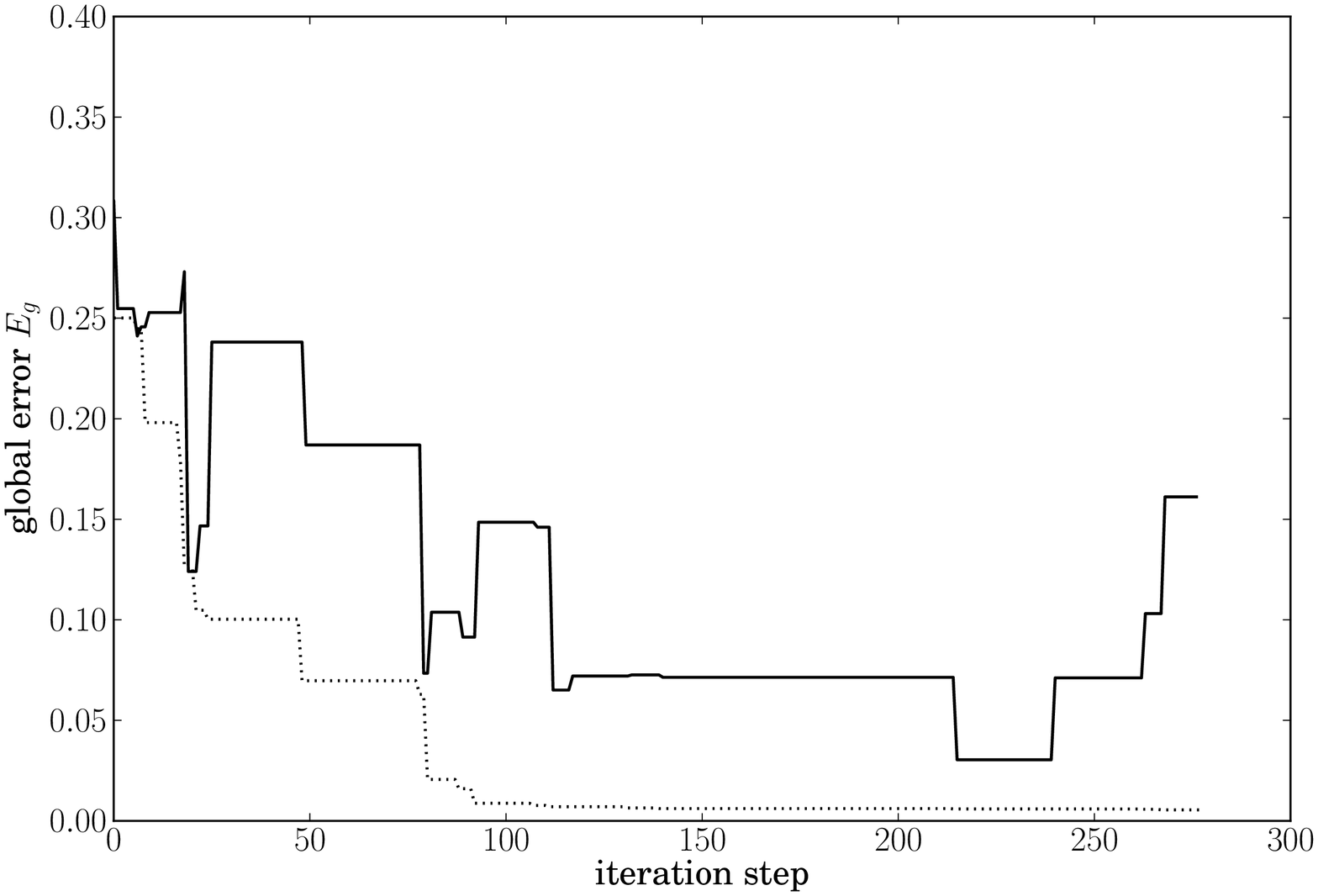}
    \includegraphics[width=0.9\columnwidth]{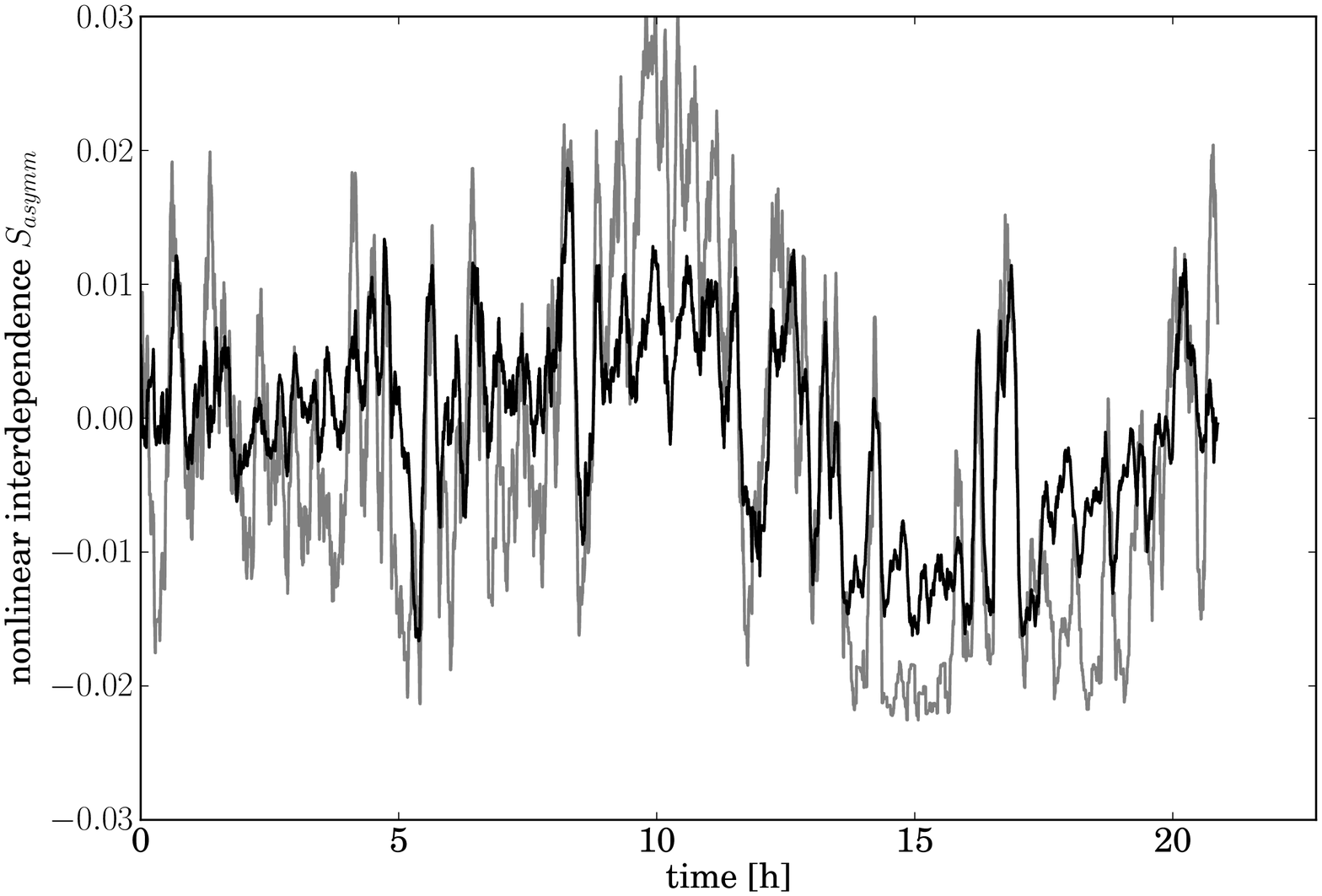}
    \caption{Exemplary findings obtained with \CNNcan.
        Top: Global error \symbGlobFehler (cf. \eref{hd:eq:globFehler}) depending on the number of iteration steps during in-sample optimization (dotted line) and during out-of-sample validation (solid line).
        Bottom: Temporal evolution of analytically calculated (black line) and approximated nonlinear interdependence $S_\text{asymm}$ (gray line) between the seizure generating area and its immediate surrounding.
        Profiles are smoothed using a 15-point ($\approx$ \unit[5]{min}) moving-average filter for better visualization.
        The CNN-approximation was performed with the templates \symbKopplungFB and \symbKopplungFF and the global cell bias~\symbBias obtained in iteration step 80.}
    \label{hd:fig:resstd}
\end{figure}

\begin{figure}[ht]
    \centering
    \includegraphics[width=0.9\columnwidth]{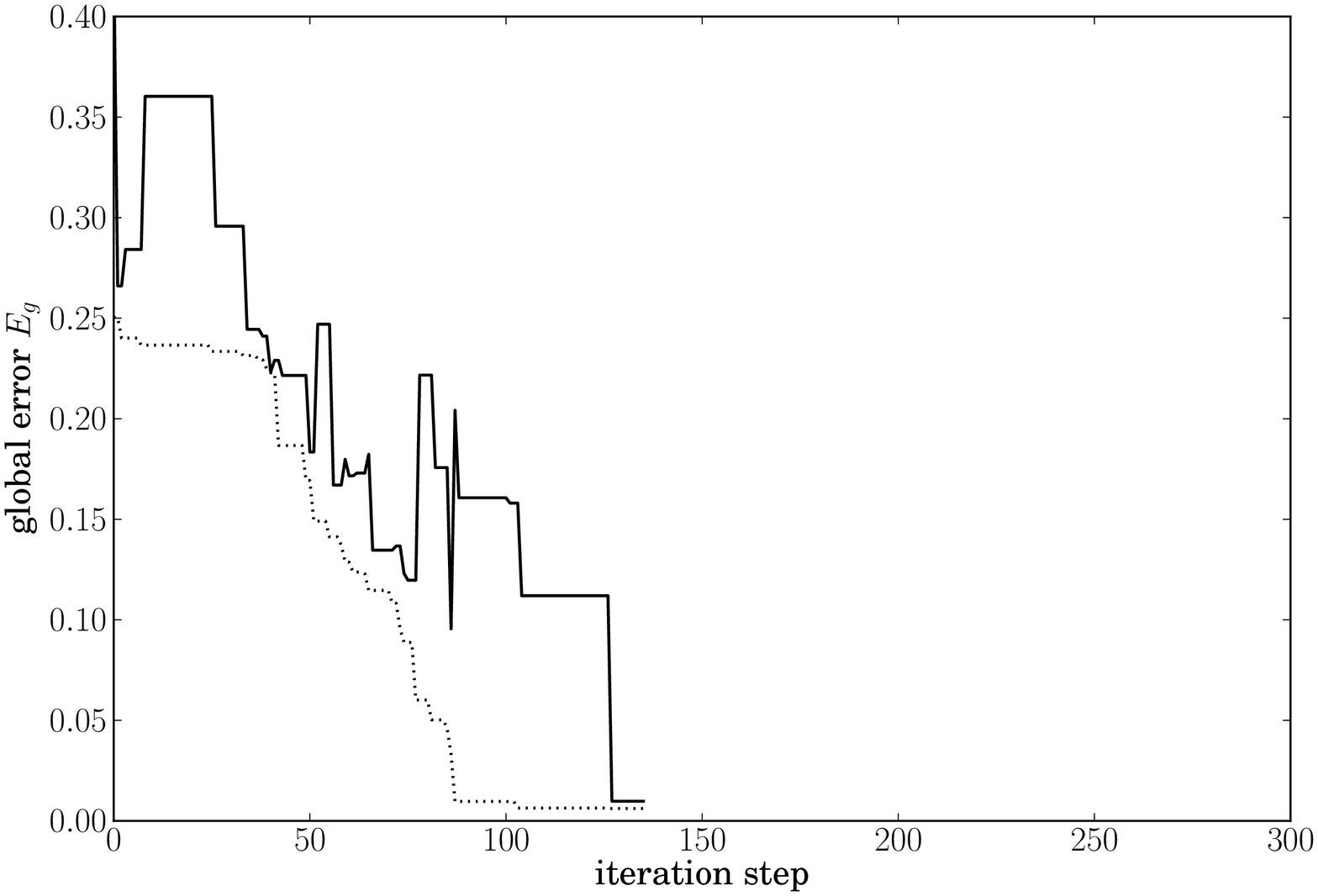}
    \includegraphics[width=0.9\columnwidth]{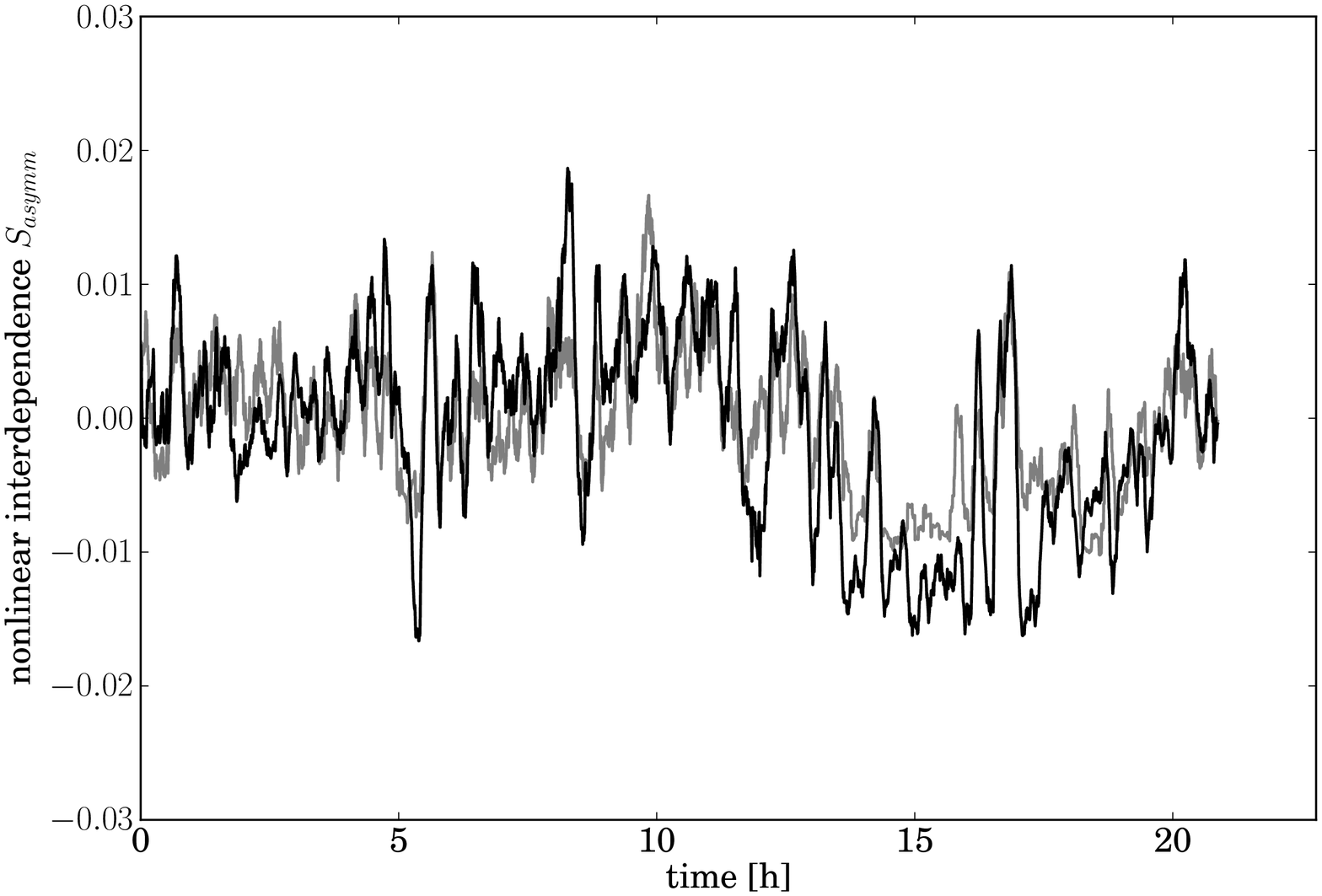}
    \caption{Same as \fref{hd:fig:resstd} but for \CNNlr.
        The CNN-approximation of the nonlinear interdependence $S_\text{asymm}$ was performed with the templates \symbKopplungFB and \symbKopplungFF and the global cell bias~\symbBias obtained in iteration step 58.}
    \label{hd:fig:resakf}
\end{figure}

\begin{figure}[ht]
    \centering
    \includegraphics[width=0.9\columnwidth]{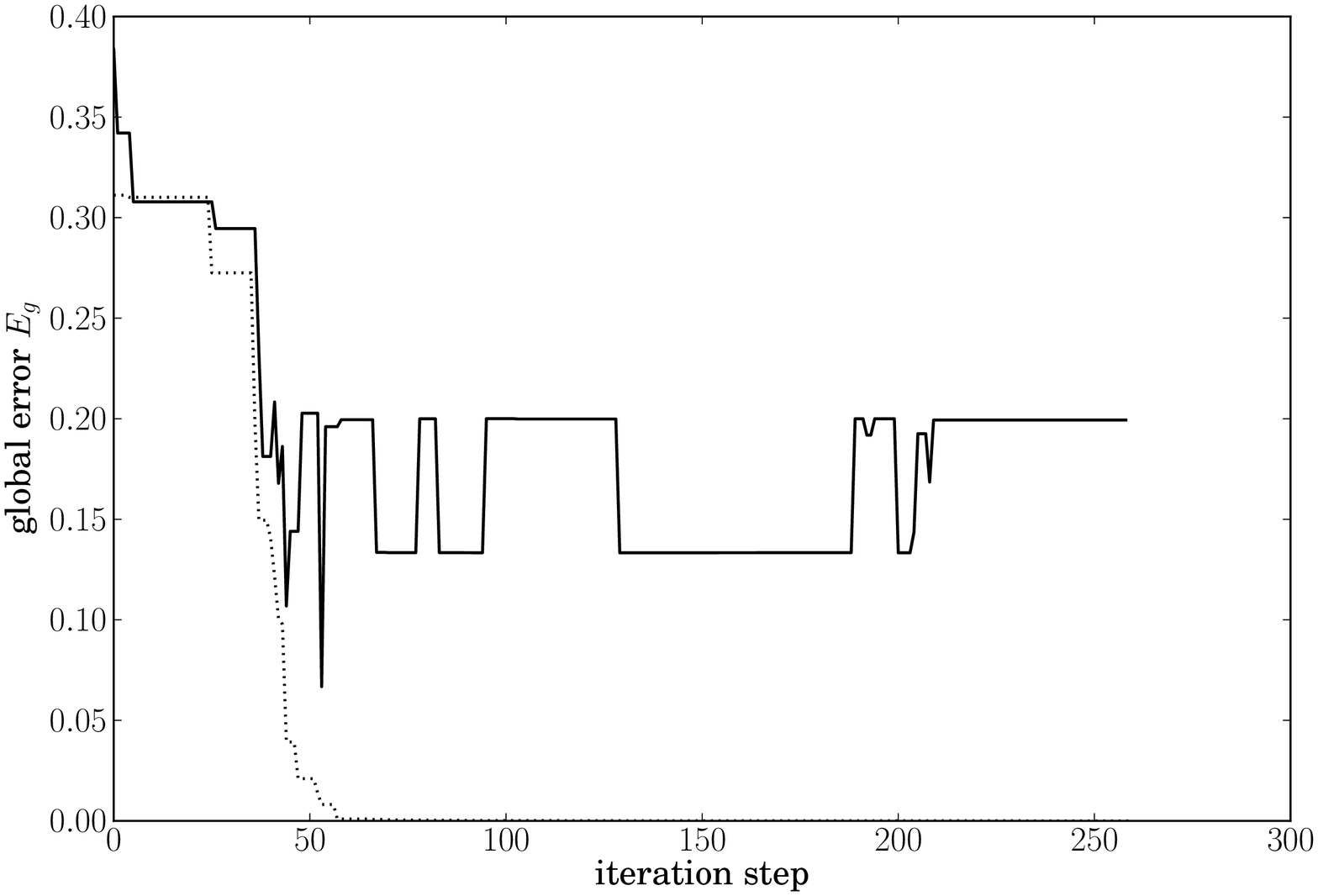}
    \includegraphics[width=0.9\columnwidth]{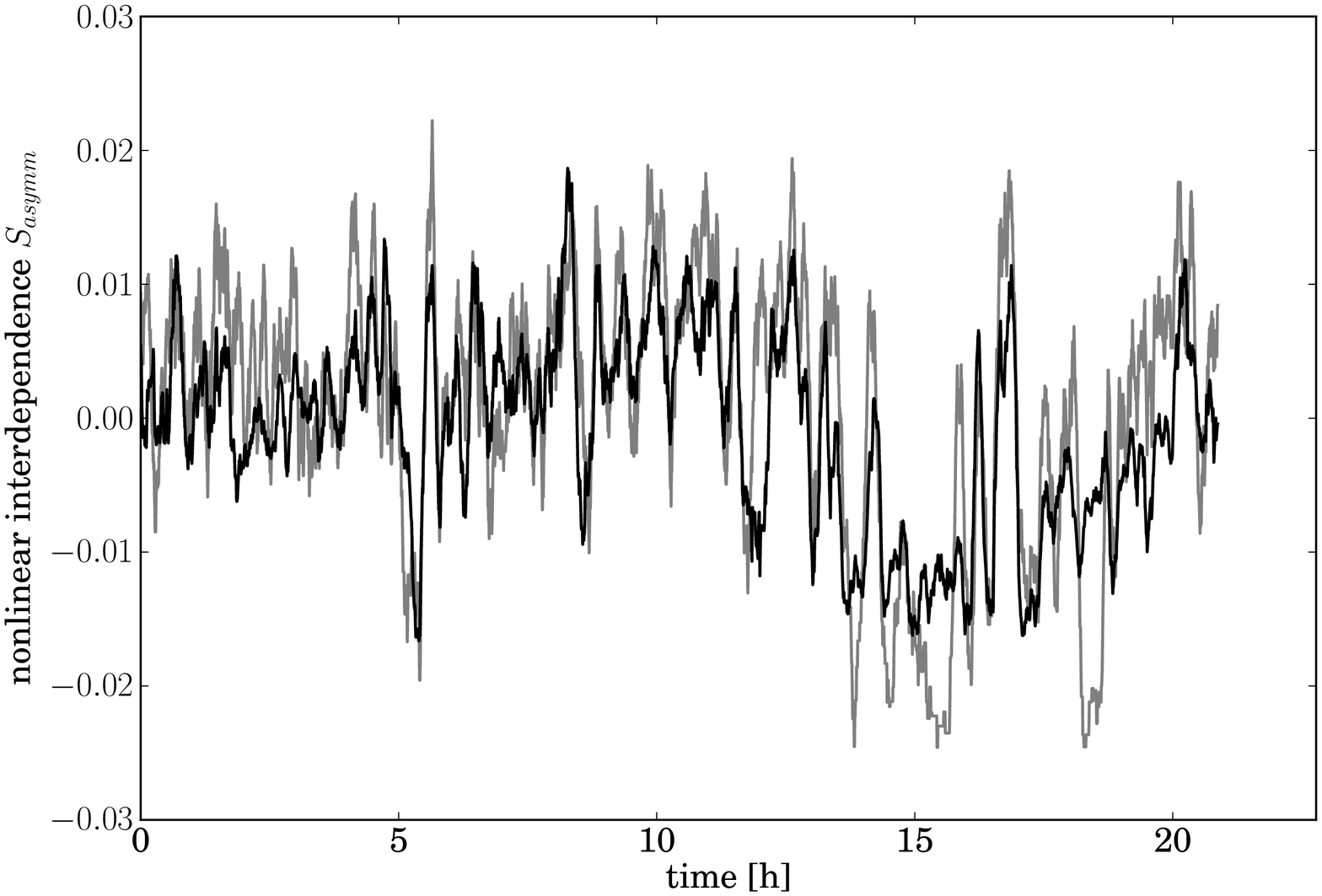}
    \caption{Same as \fref{hd:fig:resstd} but for \CNNran.
        The CNN-approximation of the nonlinear interdependence $S_\text{asymm}$ was performed with the templates \symbKopplungFB and \symbKopplungFF and the global cell bias~\symbBias obtained in iteration step 40.}
    \label{hd:fig:resrand}
\end{figure}

First of all we note that short- and long-term fluctuations of nonlinear interdependencies between brain regions could well be approximated even with CNN with complex connection topologies.
This observation does not only extend previous findings\cite{Krug2007} but it also indicates that the use of complex topologies leads to a faster optimization and validation of the CNN (cf. upper parts of \fref{hd:fig:resstd}--\ref{hd:fig:resrand}).
More importantly, when comparing performance data of the three investigated CNN, approximation accuracy increased from \unit[87.5]{\%} (\CNNcan) to \unit[90.1]{\%} (\CNNran) to \unit[92.7]{\%} (\CNNlr) (cf. lower parts of \fref{hd:fig:resstd}--\ref{hd:fig:resrand}).

\section{Conclusion}
We have investigated whether a CNN-based approximation of directed interactions between the dynamics of different areas of the human epileptic brain can be improved by replacing the
lattice-like arrangement of CNN cells by complex connection topologies.
Findings obtained from an exemplary analysis of directed interactions in intracranial electroencephalographic recordings from an epilepsy patient indicate that complex connection topologies allow for a faster optimization of CNN together with an improved
approximation accuracy of nonlinear interdependence. Our preliminary though promising findings need to be validated on the data from a larger group of patients.

Although our findings are, at present, restricted to simulated or digital realizations of
CNN, their powerful computational capacity and generalization capability combined with small size and low power consumption of hardware realizations render these networks highly attractive for the development of miniaturized seizure prediction devices.
\section*{Acknowledgments}
This work was supported by the Deutsche Forschungsgemeinschaft (Grand No. LE660/2-4).

\bibliographystyle{ws-procs9x6}

\end{document}